\documentclass[aps,prl,showpacs,twocolumn]{revtex4}
\usepackage{amsmath,bm}
\usepackage[dvips]{graphicx}

\begin{document}

\title{Acceleration of chemical reaction by chaotic mixing}
\author{M. Chertkov$^a$ and V. Lebedev$^{a,b}$}
\affiliation{$^a$ Theoretical Division, LANL, Los Alamos, NM 87545, USA \\
$^b$ Landau Institute for Theoretical Physics, Moscow, Kosygina 2,
117334, Russia}
\date{January 10, 2003}

\begin{abstract}

  Theory of fast binary chemical reaction, ${\cal A}+{\cal
    B}\to{\cal C}$, in a statistically stationary chaotic flow at
  large Schmidt number $\mbox{Sc}$ and large Damk\"ohler number
  $\mbox{Da}$ is developed. For stoichiometric condition we identify
  subsequent stages of the chemical reaction. The first stage
  corresponds to the exponential decay, $\propto\exp(-\lambda t)$
  (where $\lambda$ is the Lyapunov exponent of the flow), of the
  chemicals in the bulk part of the flow. The second and the third
  stages are related to the chemicals remaining in the boundary
  region. During the second stage the amounts of ${\cal A}$ and ${\cal
    B}$ decay $\propto 1/\sqrt{t}$, whereas the decay law during
  the third stage is exponential, $\propto\exp(-\gamma t)$, where
  $\gamma\sim\lambda/\sqrt{Sc}$.

%LAUR 021743


\end{abstract}

\maketitle

A common expectation is that random advection should essentially
accelerate chemical reactions in fluid phase, since it should lead to
homogenization of the reaction mixtures. Then dynamics is determined by an
interplay of three factors: diffusion, advection and chemical reaction
rate. Typical situation realizing in chemical reactors is that
chemical reaction itself is much faster than mixing and diffusion,
i.e. the Damk\"{o}hler number, $\mbox{Da}$, which is defined as the
ratio of the mixing time to the characteristic time of the reaction
\cite{36Dam}, is large. For the binary reaction this separation of
temporal scales results in formation of lamellar structure, build of
stripes, populated solely by one chemical. The stripes of different
chemicals are separated from each other by an interface of complicated
shape, and the chemicals co-exist only in the narrow interface domain
where the chemical reaction occurs. The reaction is limited by
diffusion in the sense that diffusion controls fluxes of the chemicals
into the interfacial reaction zone \cite{63FK,76Hil,84CO}. The
physical picture of the acceleration due to the random advection is
that it stretches domains populated by one chemical into thin sheets,
so that the chemical reaction driven by diffusion proceeds more
efficiently because of an essential increase of the interface area. In
this letter we explain how this general physical picture, formulated
initially for unbounded flows
%(see, e.g., \cite{84CO}),
applies to chaotic flows confined to a finite geometry.

We consider a binary chemical reaction, ${\cal A}+{\cal B}\to{\cal
  C}$, in a dilute solution of two chemicals. We study decay problem,
with an initial distribution of the chemicals ${\cal A}$ and ${\cal
  B}$, created by injecting solution of one chemical, say of ${\cal
  A}$, into solution of the other chemical, ${\cal B}$. It is assumed
that the inverse reaction ${\cal C}\to{\cal A}+{\cal B}$ is
negligible, i.e. there is no back influence of ${\cal C}$ on the
distribution of ${\cal A}$ and ${\cal B}$. Then molecular
concentrations of the chemicals, $n_a$ and $n_b$, vary according to
the following non-linear governing equations (see, e.g.,
\cite{Landau})
 \begin{eqnarray}
 \partial_t n_{a,b}+({\bm v}\cdot\nabla)n_{a,b}
 =\kappa_{a,b}\nabla^2 n_{a,b}-R n_a n_b \,,
 \label{nab} \end{eqnarray}
where $R$ is the reaction rate coefficient, ${\bm v}$ is the velocity
of the flow, which is assumed to be incompressible ($\nabla\cdot\bm
v=0$), $\kappa_{a,b}$ are the diffusion coefficients of the chemicals.
Here, we assume that the fluid dynamics is independent of the chemical
reaction, that is the velocity field does not sense changes in the
chemical concentrations nor heat is released in the result of the
chemical reaction. Our approach is also applicable to the case,
realized in tubular chemical reactors, when the solution of the
chemicals, prepared at the entrance, is then pushed through a pipe. In
this case the position along the pipe plays the role of time in the
decay problem.

The major question addressed in the letter is: how do the total
amounts of chemicals, $N_{a,b}=\int\mbox d{\bm r}\,n_{a,b}(t,{\bm
  r})$, decay as time $t$ advances? We focus primarily on the
stoichiometric case $N_a=N_b$. This case is of major interest for
applications, as it allows to get pure product ${\cal C}$ (not mixed
with the reagents), by the time reaction is completed. (An effect of
a mismatch between $N_a$ and $N_b$ is briefly discussed later in this
text.) We identify major stages of the chemical reaction and relate
them to the chemicals decay in different parts of the flow. An essential
part of the evolution is related to the boundary region.

We discuss mainly the case of $\kappa_a=\kappa_b =\kappa$. (It is
argued later in the text that $\kappa_a\neq \kappa_b$ does not lead to
significant changes in the theory.) Then one obtains a closed equation
for the difference field, $n=n_a-n_b$,
 \begin{eqnarray}
 \partial_t n + (\bm v\cdot\nabla)n
 =\kappa \nabla^2 n \,,
 \label{neq} \end{eqnarray}
from Eq. (\ref{nab}), i.e. one finds that $n(t,{\bm r})$ is a passive
scalar field. Note, that $n$ has no definite sign, and that $\int\mbox
d\bm r\, n=0$ in the case of perfect matching, $N_a=N_b$.

We assume that the chaotic statistically steady velocity field ${\bm
  v}(t,\bm r)$ contains only few harmonics of the reservoir size, i.e.
the flow is smooth. This regime can be realized in chemical reactors
with mechanically rotating mixers or externally driven magnets
stirring the fluid in the perfect mixing devices and also in the
tubular reactors at moderate Reynolds numbers. (See \cite{99BB} for a
discussion of the chemical engineering principles behind various
reactor designs.) Complementary to its practical significance, passive
scalar advection in a smooth chaotic flow is also a well studied (by
both theoretical \cite{59Bat,67Kra,94SS,95CFKLa,99BF} and experimental
\cite{97WMG,00JCT,01GSb} means) subfield of statistical hydrodynamics.
(See also reviews \cite{00SS,01FGV}.) The passive scalar decay theory,
developed in \cite{99Son,99BF} for an unbounded flow, was recently
modified for bounded flows, i.e. for chaotic flows with suitable (no
slip) conditions on the boundary \cite{02CLa}. Smoothness of the flow
allows one to approximate the velocity difference between close points
by a linear, although fluctuating in time, profile. In the bulk
region, the linear profile approximation is valid for separations
smaller than the system size $L$. In the periphery, i.e. close to the
solid boundary (wall), the linear profile approximation is valid for
velocity fluctuations on a scale smaller than distance to the
boundary. An important consequence of the linear velocity profile
approximation is that close Lagrangian trajectories diverge
exponentially in time. The mean logarithmic rate of the nearby
Lagrangian trajectories divergence defines the Lyapunov exponent of
the flow, $\lambda$. Notice, that in the peripheral domain advection
is essentially anisotropic, and the stretching rate along the boundary
is estimated by $\lambda$, while the stretching rate in the direction
normal to the boundary is significantly smaller.

\begin{figure}[tbp]
 \includegraphics[width=6.5cm,height=5.5cm]{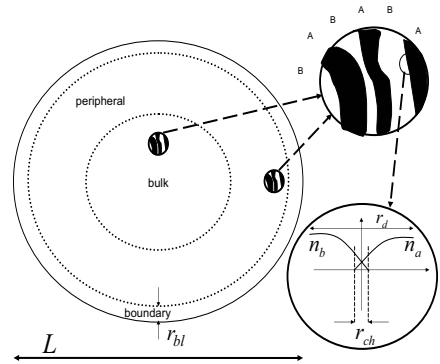}
 \caption{Chemical reactor boundary is drawn by solid line on the
 main chart. Dotted lines separate bulk, peripheral and boundary
 domains of the flow. Chart in the upper right corner shows schematically
 the stripe structure magnified from the bulk and/or peripheral
 domains. Black and white regions are the ones populated by ${\cal A}$
 and ${\cal B}$ respectively. The chart in the low right corner shows
 (under even stronger magnified glass) distribution of chemicals
 normal to their contact interface.}
 \label{One} \end{figure}

We now discuss characteristic spatial scales in the problem. The size
of the system, $L$, which is also the chaotic flow typical eddy scale,
is the largest scale in the problem. A comparison of the advection and
diffusion terms in Eq. (\ref{nab}) sets the dissipative scale of the
flow, which in the bulk region is $r_d=\sqrt{\kappa/\lambda}$. We
assume that the Schmidt number, $\mbox{Sc}\sim(L/r_d)^2$, is large,
i.e. in the asymptotically wide range of scales, $L\gg r\gg r_d$,
advection dominates diffusion. The width of the diffusive boundary
layer is estimated by $r_{bl}\sim \mbox{Sc}^{1/4}r_d$, i.e.
$r_{bl}>r_d$. Yet another important scale, associated with the
chemical reaction itself, is the size of the reaction zone, $r_{ch}$
(the width of the interfacial domain where the chemical reaction
occurs). In the bulk region the scale is estimated by $r_{ch}=r_d
[\lambda/(Rn_{m})]^{1/3}$, where $n_{m}$ is a typical concentration of
the chemicals inside the layers. (The estimate for the width of the
reaction zone should be modified near the boundary, where it appears
to be larger than in the bulk.) Initially, $r_{ch}$ is much smaller
than $r_d$ (and, consequently, than $L$); the inequality is a
consequence of the $\mbox{Da}\gg 1$ assumption. (Indeed, in accordance
with the definition, the Damk\"ohler number can be estimated as
$\mbox{Da}\sim R n_0/\lambda$, where $n_0$ is a typical value of the
initial chemical concentration. Thus, at $t=0$, $r_{ch}\sim
\mbox{Da}^{-1/3}r_d$.) However, $r_{ch}$ grows as $n_{m}$ decreases.
Thus even though the separation of scales is perfect initially, it
eventually breaks down at the latest stage of the chemical reaction. A
cartoon illustration of the scale hierarchy is shown in Fig.
\ref{One}. The magnified striped structure is shown on the chart in
the upper right corner of the figure. Regions populated by one
chemical are single-colored. To resolve the interface domain, even
stronger magnification is needed. Dependence of the chemicals
concentrations on the coordinate normal to the interface is shown
schematically on the chart in the lower right corner of the figure.

The separation of scales, $r_{ch}\ll r_d, L$, allows an important
simplification in the description. Indeed, the chemical reaction takes
place in the $r_{ch}$-narrow interface domain, where the values of
$n_a$ and $n_b$ are comparable. Outside this narrow region, {i.e.}  in
the region dominated by one of the chemicals, the presence of the
other chemical is negligible. Thus, in the limit $r_{ch}\to0$, {i.e.}
when the reaction zone becomes infinitesimally thin, one obtains
$n_a=n, \ n_b=0$ for $n>0$ and $n_a=0, \ n_b=-n$ for $n<0$. These
relations imply a remarkable conclusion \cite{63FK,84CO}: the fast
chemical reaction can be described in terms of the linear setting
(\ref{neq}) which does not contain the chemical reaction rate
coefficient $R$. The reaction rate is determined by the diffusion
fluxes of ${\cal A}$ and ${\cal B}$ to the $n=0$ surface. These fluxes
are equal to each other and opposite in sign, which is translated, at
$r_{ch}\to 0$, into a continuity condition for $\nabla n$ at the
interface. This observation also means that, while $r_{ch}$ is much
smaller than all other relevant scales, our problem is reduced to the
problem of scalar decay in chaotic bounded flow.

Furthermore, from Eq. (\ref{neq}), it is straightforward to derive
equations for correlation functions of $n$ (the derivation procedure
is similar to the one described in \cite{95CFKLa,99BF}). The equation
for the mean value of $n$, $\langle n\rangle$, derived by averaging
over times larger than the correlation time of the flow $\tau_0$, is
 \begin{eqnarray}
 \partial_t\langle n\rangle=\nabla_\alpha
 (D_{\alpha\beta}\nabla_\beta\langle n\rangle)
 +\kappa\nabla^2\langle n\rangle\,.
 \label{nnn} \end{eqnarray}
In the case of a short-correlated (in time) flow, $\lambda \tau_0\ll
1$, the turbulent diffusion tensor $D$ is expressed through the
velocity pair correlation function: $D_{\alpha\beta}(\bm
r)=\int_0^\infty\mbox dt\, \langle v_\alpha(t,\bm r) v_\beta(0,\bm
r)\rangle$. Eq. (\ref{nnn}) remains valid in the general, not
necessarily short-correlated limit. However, the relation between the
eddy-diffusivity tensor and the velocity correlations becomes more
complicated. $D(\bm r)$ tends to zero when $\bm r$ approaches the
boundary, since the flow velocity tends to zero there. (The
longitudinal component $D_\|$ of the tensor ${D}$ behaves as
$D_\|\propto q^2$, whereas its transverse component $D_\perp$ behaves
as $D_\perp\propto q^4$, where $q$ is the distance to the boundary.)
Our description of the chemical reaction problem is based on the
solutions of Eq. (\ref{nnn}) in different spatio-temporal domains.
Knowing $\langle n\rangle$, one can establish the temporal behavior of
$N_{a,b}$. The brief style of this letter does not allow us to present
the complete analysis here. Therefore, below we report final results,
omitting details of the derivation. To clarify the results, we also
pay special attention to presenting a physical picture of the
phenomenon.

We find that the chemical reaction (which starts at $t=0$)
undergoes the following four stages:

I. {\em Formation of stripes in the bulk.} Advection creates from an
initially smooth distribution a striped structure of alternating
domains of ${\cal A}$ and ${\cal B}$ \cite{84CO}. The stripes become
dynamically thinner, {i.e.} inhomogeneities of smaller and smaller
scales are produced.  Once the width of the stripe decreases down to
the diffusive scale, $r_d$, the stripe collapses (wiped out by the
diffusion-limited chemical reaction) in a time $\sim\lambda^{-1}$.
Since the stretching (contraction) process leading to creation of the
stripes is exponential in time
\cite{59Bat,67Kra,94SS,95CFKLa,00SS,01FGV}, the initial stage (when
the $r_d$-stripes are formed) lasts for $\tau_1\sim\ln
(\mbox{Sc})/\lambda$, i.e. just the time required for the cascade of
passive scalar to run from $L$ down scale to $r_d$. Even though the
interfacial area increases exponentially during the first stage,
$N_{a,b}$ do not vary significantly. By the end of this stage the bulk
parts of $N_{a,b}$ begin to decay rapidly (exponentially), with a
decrement of the order of $\lambda$, according to the law of the
passive scalar decay in an unbounded spatially smooth flow
\cite{99Son,99BF}.
%This exponential decay is in agreement with
%numerical study of fast chemical reaction in a periodic chaotic flow
%\cite{96ML}.
Thus, after the first stage the chemicals remain mainly
in the peripheral region.

Notice also, that after the first stage, stripes of different widths, distributed between $r_d$
and $L$, are present in the bulk. (This multi-scale structure is also seen in the passive scalar
decay experiment \cite{00JCT,01GSb}, and in binary-reaction numerics \cite{96ML}.) When the
$r_d$-wide stripe, say, of the chemical ${\cal A}$ collapses, then two nearby stripes of the
chemical ${\cal B}$ form one wider stripe. Thus, collapse of $r_d$-narrow stripes is accompanied
by creation of wider stripes, which are shrunk by the flow in turn, and so on and so forth.

II. {\em Peripheral-region-dominated dynamics.} The same process of
layered structure formation takes place in the peripheral domain as
well. However, advection, which is statistically isotropic in the
bulk, is strongly anisotropic in the peripheral domain, where
advection is more efficient in the direction along the boundary than
in the normal direction. This anisotropy causes the layers in the
peripheral domain to stretch mainly along the boundary. The stripes
closer to the boundary shrink slower than the remote ones, since the
normal to the boundary component of the stretching rate decreases as
one approaches the boundary. Therefore the developed layered structure
(i.e. the one which contains stripes of the diffusive scale width)
occupies a part of the peripheral region where the amounts of $\cal A$
and $\cal B$ become negligible. Thus, the empty of chemicals region,
formed in the bulk by the end of the first stage, starts to expand
towards the boundary. As a result, the chemicals are arranged mainly
within a $\delta$-vicinity of the boundary (wall), $\delta\sim
L/\sqrt{\lambda t}$, where the concentrations of the chemicals remain
practically unchanged. Outside this layer, at $L\gg q\gg\delta$ (where
$q$ is the separation from the boundary), the concentration of
chemicals decreases algebraically $\langle n_{a,b}\rangle\propto
t^{-3/2}q^{-3}$. During this stage the overall amounts of chemicals
decrease as $\delta(t)$, that is $\propto1/\sqrt t$. The
spatio-temporal picture explained above follows from the universal
form of the velocity field profile in the proximity of the boundary.
This stage lasts for $\tau_2\sim \sqrt{Sc}/\lambda$, i.e. until
$\delta$ shrinks to the width of the boundary layer, $r_{bl}$.

III. {\em Boundary-layer-dominated dynamics.} Chemicals remain mainly
within the $r_{bl}$-thin (not varying with time) vicinity of the
boundary. The boundary layer width, $r_{bl}$, is still much larger
than the reaction zone size (defined for the boundary region), so that
the passive scalar description applies. The interfacial area where the
chemicals interact does not change significantly anymore. Thus, due to
linear relation between flux of chemicals to the interface and their
concentrations, the algebraic decay switches to an exponential one,
{i.e.} $\langle n_{a,b}\rangle \propto \exp(-\gamma t)$, for
$t\gg\tau_2$, where $\gamma\sim\lambda /\sqrt{Sc}\sim
L^{-1}\sqrt{\lambda \kappa}$. (Note, also, that the slow-exponential
regime, derived in \cite{02CLa} for the passive scalar, is consistent
with the experimental observations of \cite{01GSb}.) Then,
$N_{a,b}(t)\propto\exp(-\gamma t)$. Chemicals are mainly concentrated
inside the diffusion boundary layer. Outside the boundary layer (at
$q\gg r_{bl}$) one of the chemicals prevails and its concentration
decays algebraically, $\propto 1/q^3$. The passive scalar description
in the vicinity of the boundary layer is broken when $r_{ch}$, which
grows exponentially with time, becomes of the order of $r_{bl}$,
{i.e.} when at the boundary $n_{a,b}$ becomes
$\sim\lambda/(R\sqrt{\mbox{Sc}})$. One concludes that the duration of
the boundary-layer-dominated stage is $\tau_3\sim\gamma^{-1} \ln(R
n_0\sqrt{\mbox{Sc}}/\lambda)$, where $n_0$ is the initial
concentration of the chemicals.

IV. {\em Nonlinear stage.} By the end of the previous stage, advection
and diffusion homogenize the remaining amounts of the chemicals,
first, within the boundary layer and later over the entire reservoir.
After that there are no inhomogeneities of $n_{a,b}$ left in the
system. A purely homogeneous kinetic process takes over: $\mbox
dN_{a,b}/\mbox dt=-R V N_aN_b$ (where $V$ is the chemical reactor
volume). Thus, $N_{a,b}\propto 1/t$, during the final stage.

If $N_a\neq N_b$ then the proposed scheme is valid until $N_{a,b}$
become of the order of $|N_a-N_b|$. Then $N_a$ saturates to a constant
(if $N_a>N_b$) and $N_b$ disappears exponentially, $\propto\exp(-R N_a
t/V)$. (Note that the exponential decay starts after a short
intermediate stage characterized by complete homogenization of ${\cal
A}$ due to advection and diffusion.)

Let us now discuss the effect of unequal diffusion coefficients, still
assuming that $\sqrt{\kappa_{a,b}/\lambda}\ll L$. If $\mbox{Da}\gg1$
then during the first stages, the chemical length $r_{ch}$ is (as
above) much smaller than all other scales. This problem can also be
reduced to a linear one considering the advection-diffusion equations
in domains populated by different species, supplemented by the
condition that fluxes of the two chemicals towards the interface are
equal. During the first two stages, the evolution is controlled by the
stripe formation process which is insensitive to the diffusion. During
the latter, third and fourth, stages of the evolution in the uneven
$\kappa_a\sim \kappa_b$ case the chemicals evolve similarly to what
was described above for the, $\kappa_a=\kappa_b$, case. Thus, the
above description applies to the general, $\kappa_a\neq\kappa_b$, case
as well.

We conclude with some general remarks. A complicated spatio-temporal
behavior for the binary chemical reaction in a chaotic flow is
established. Evolution of the chemicals near the boundary (where
mixing is slower than in the bulk) determines the intermediate stages
of the reaction. Those boundary-dominated stages were not singled out
in the previous publications on the subject simply because the stages
are not observed in an infinite \cite{84CO} or periodic \cite{96ML}
flow systems. In our setting the lamellar structure (which is
statistically isotropic in bulk and strongly anisotropic near
boundaries) is dynamically generated by advection. (This situation is
essentially different from the one considered in \cite{90MO}, where
the lamellar structure is created initially and no advection
participates in subsequent evolution.) We focused here on large scale
chaotic flows with the size of the box being of the order of the major
scale of the flow. However, it is also of interest for applications to
describe chemical reaction acceleration in turbulent flows, which are
smooth only inside the viscous range of scales. In this case with a
large value of the viscous to dissipative scales ratio, a
consideration similar to those presented in the letter is applicable.
We plan to examine the more complicated case in the future.  Also, the
approach developed in the paper is generalizable for other, more
complicated, types of chemical reaction, e.g. competing chemical
reactions. For completeness, let us also mention another case of
interest which is realized at moderate $\mbox{Da}$, large $\mbox{Sc}$
and if one of chemicals is present in abundance. The joint effect of
advection and chemistry is different in this case (than in the problem
discussed in this letter), even though rich multi-scale structure of
spatial correlations is also revealed \cite{98Che}. A final remark
concerns the validity of the hydrodynamic description of the chemical
reaction dynamics. It is known that the character of spatial
fluctuations in the initial distribution of chemicals may essentially
influence the long-time behavior in diffusion-limited chemical systems
\cite{78OZ,83TW}. In some cases (of low space dimensionality, $d\leq
2$) large scale renormalization of the concentration fields due to the
small scale fluctuations could be important. (See, e.g.,
\cite{95Cara}.) In our case, however, this does not happen because the
long-time correlations are completely destroyed by chaotic advection.

We thank J. Cardy, G. D. Doolen, A. Fouxon, A. Groisman, I. Kolokolov,
V. Steinberg, P. Tabeling, and Z. Toroczkai for helpful discussions
and comments.


\begin{thebibliography}{99}

%\bibitem{90Eps}
%I. R. Epstein, Nature (London) {\bf 346}, 16 (1990).

%\bibitem{97PT}
%O. Paireau and P. Tabeling, Phys. Rev. E {\bf 56}, 2287 (1997).

\bibitem{36Dam}
G. Z. Damk\"{o}hler, Electrochem. {\bf 42}, 846 (1936).

\bibitem{63FK}
S. K. Friedlander and K. H. Keller, Chem. Engn. Sci {\bf 18}, 365 (1963).

\bibitem{76Hil}
J. C. Hill, Ann. Rev. Fluid. Mech {\bf 8}, 135 (1976).

\bibitem{84CO}
R. Chella and J. M. Ottino, Chem. Eng. Sci. {\bf 39}, 551 (1984).

\bibitem{Landau}
L. D. Landau and E. M. Lifshitz, {\it Statistical Physics, part
1}, (Pergamon Press, 1980).

\bibitem{99BB}
J. Baldyga and J. R. Bourne, Turbulent mixing and chemical reactions,
John Wiley \& Sons, 1999.

\bibitem{59Bat}
G. K. Batchelor, JFM {\bf 5}, 113 (1959).

\bibitem{67Kra}
R. H. Kraichnan, Phys. Fluids. {\bf 10}, 1417 (1967), JFM {\bf
47}, 525 (1971); Ibid {\bf 67}, 155 (1975).

\bibitem{94SS}
B. I. Shraiman and E. D. Siggia, Phys. Rev. E {\bf 49}, 2912
(1994).

\bibitem{95CFKLa}
M. Chertkov, G. Falkovich, I. Kolokolov, and V. Lebedev, Phys.
Rev. E {\bf 51}, 5609 (1995).

\bibitem{99BF}
E. Balkovsky and A. Fouxon, Phys. Rev. E {\bf 60}, 4164 (1999).

\bibitem{97WMG}
B. S. Williams, D. Marteau, and J. P. Gollub, Phys. Fluids {\bf
9}, 2061 (1997).

\bibitem{00JCT}
M.-C. Jullien, P. Castiglione, and P. Tabeling, Phys. Rev. Lett.
{\bf 85}, 3636 (2000).

\bibitem{01GSb}
A. Groisman and V. Steinberg, Nature {\bf 410}, 905 (2001).

\bibitem{00SS}
B. I. Shraiman and E. D. Siggia, Nature (London) {\bf 405}, 639
(2000).

\bibitem{01FGV}
G. Falkovich, K. Gaw\c{e}dzki, and M. Vergassola, Rev. Mod. Phys.
{\bf 73}, 913 (2001).

\bibitem{99Son}
D. T. Son, Phys. Rev. E {\bf 59}, R3811 (1999).

\bibitem{02CLa}
M. Chertkov and V. Lebedev, Phys. Rev. Lett. to appear 02/2003.

\bibitem{96ML}
F. J. Muzzio and M. Liu, Chem. Engn. J {\bf 64}, 117 (1996).

\bibitem{90MO}
F. J. Muzzio and J. M. Ottino, Phys.Rev.A {\bf 40}, 7182 (1989); {\bf 42}, 5873 (1990).

\bibitem{98Che}
M. Chertkov, Phys. Fluids {\bf 10}, 3017 (1998); {\bf 11}, 2257
(1999).

\bibitem{78OZ}
A. A. Ovchinnikov and Ya. B. Zeldovich, Chem. Phys. {\bf 28}, 215
(1978).

\bibitem{83TW}
D. Toussaint and F. Wilczek, J. Chem. Phys. {\bf 78}, 2642 (1983).

\bibitem{95Cara}
J. Cardy, J. Phys. A: Math. Gen. {\bf 28}, L19 (1995).

%\bibitem{Batchelor}
%G. K. Batchelor, {\it An Introduction to Fluid Dynamics},
%Cambridge University Press, 1967.

\end{thebibliography}
\end{document}